\documentclass[preprint,showpacs,preprintnumbers,amsmath,amssymb,
superscriptaddress]{revtex4}

\usepackage{graphicx}
\usepackage{dcolumn}
\usepackage{bm}

\begin{document}


\title{DOUBLE BETA DECAY EXPERIMENTS: BEGINNING OF A NEW ERA}

\author{A.S.~Barabash} \email{barabash@itep.ru}
\affiliation{Institute of Theoretical and Experimental Physics, B.\
Cheremushkinskaya 25, 117218 Moscow, Russia}


\begin{abstract}
The review of current experiments on search and studying of double beta
decay processes is done. Results of the most sensitive experiments are
discussed and values of modern limits on effective Majorana neutrino mass
($\langle m_{\nu}\rangle$) are given. New results on two neutrino double beta 
decay are presented. 
The special attention is given to new current experiments with mass of studied 
isotopes more than 100 kg, EXO--200 and KamLAND--Zen. These experiments open
a new era in research of double beta decay. In the second part of the review
prospects of search for neutrinoless double beta decay in new experiments with
sensitivity to $\langle m_{\nu}\rangle$ at the level of $\sim 0.01-0.1$ eV are 
discussed. Parameters and
characteristics of the most perspective projects (CUORE, GERDA, MAJORANA, 
SuperNEMO, EXO, KamLAND--Zen, SNO+) are given.
\end{abstract}



\maketitle

\section{Introduction}

Interest in $0\nu\beta\beta$ decay has seen a significant renewal in recent 
10 years after evidence for neutrino oscillations was obtained 
from the results of atmospheric, solar, reactor, and accelerator 
neutrino experiments. These results are impressive proof that 
neutrinos have a nonzero mass. The detection and study of $0\nu\beta\beta$ 
decay may clarify the following problems of neutrino physics: 
(i) lepton number non-conservation, (ii) neutrino nature: whether 
the neutrino is a Dirac or a Majorana particle, (iii) absolute 
neutrino mass scale (a measurement or a limit on m$_1$), (iv) the 
type of neutrino mass hierarchy (normal, inverted, or quasidegenerate), 
(v) CP violation in the lepton sector (measurement of the Majorana 
CP-violating phases).

Progress in the double beta decay is connected with increase in mass 
of a studied isotope and sharp reduction of a background. During a 
long time (1948--1980) samples with mass of isotope 
$\sim$ 1--25 grams were used. So, for example, the first observation of  a two neutrino 
double beta decay in direct (counting) experiment has been done in 1987 
when studying 14 g of enriched $^{82}$Se \cite{ELL87}. And only in the 80th - 
beginning of the 90th the mass of studied isotope increased to 
hundred grams and even to 1 kg. In the 90th Heidelberg-Moscow \cite{KLA01} 
and IGEX \cite{AAL02} experiments, containing 11 kg and 6.5 kg of $^{76}$Ge, 
respectively, were started. In zero years the NEMO--3 \cite{ARN05} and 
CUORICINO \cite{AND11} installations, containing approximately 10 kg of 
isotopes ( 7 kg of $^{100}$Mo, 1 kg of $^{82}$Se, etc. in NEMO--3 and 40 kg 
of crystals from a natural oxide of Te, containing 10 kg of $^{130}$Te, 
in CUORICINO) set the fashion.

In 2011 the EXO--200 \cite{ACK11} and KamLAND--Zen \cite{GAN12} experiments 
have been started (in which 
hundreds kilograms of $^{136}$Xe are used). Soon 
it is planned to carry out start of several more experiments with 
mass of studied isotopes $\sim 100$ kg (SNO+ \cite{HAR12}, CUORE \cite{GOR12}, etc.). And it 
means the beginning of a new era in $2\beta$ decay experiments when 
sensitivity to effective Majorana mass of neutrino will reach for 
the first time values $< 0.1$ eV.

Structure of the review is the following: in section II current 
large-scale experiments on $\beta\beta$ decay are considered, in section 
III the most perspective planned experiments are discussed, the 
best modern limits on neutrino mass and the forecast for possible 
progress in the future are given in section IV (Conclusion).

\section{Current large-scale experiments}

In this section the current large-scale experiments are discussed. 
NEMO--3 experiment was stopped in January, 2011, but data analysis 
in this experiment proceeds and consequently it should be carried 
to the current experiments.

\subsection{NEMO--3 \cite{ARN05,SIM12}}

\begin{figure*}
\begin{center}
\resizebox{0.5\textwidth}{!}{\includegraphics{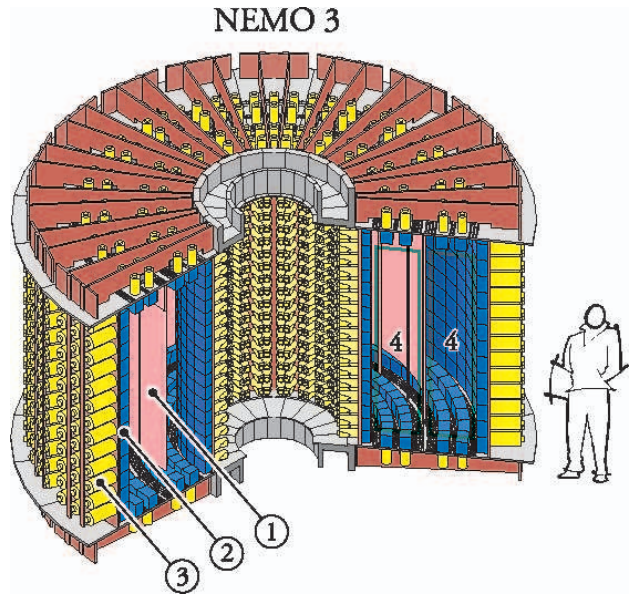}}
\caption{ The NEMO-3 detector without shielding \cite{SIM12}. 1 -- source foil;
2-- plastic scintillator; 3 -- low radioactivity PMT; 4 --
tracking chamber.}
\label{fig1}
\end{center}
\end{figure*}

This tracking experiment, in contrast to experiments with $^{76}$Ge, 
detects not only the total energy deposition, but other parameters 
of the process, including the energy of the individual electrons, 
angle between them, and the coordinates of the event in the source 
plane. Since June of 2002 and to January of 2011, the NEMO--3 detector 
has been operated in the Frejus Underground Laboratory (France) located 
at a depth of 4800 m w.e. The detector has a cylindrical structure and 
consists of 20 identical sectors (see Fig. 1). A thin (30--60 mg/cm$^2$) 
source containing double beta decaying nuclei and natural material 
foils have a total area of 20 m$^2$ and a weight of up to 10 kg was 
placed in the detector.  The energy of the electrons is measured 
by plastic scintillators (1940 individual counters), while the tracks 
are reconstructed on the basis of information obtained in the planes 
of Geiger cells (6180 cells) surrounding the source on both sides. 
The main characteristics of the detector are the following. The 
energy resolution of the scintillation counters lies in the interval 
14-17\% FWHM for electrons of energy 1 MeV. The time resolution is 
250 ps for an electron energy of 1 MeV and the accuracy in 
reconstructing the vertex of 2e$^-$ events is 1 cm.

\begin{figure}
\includegraphics[scale=1.20]{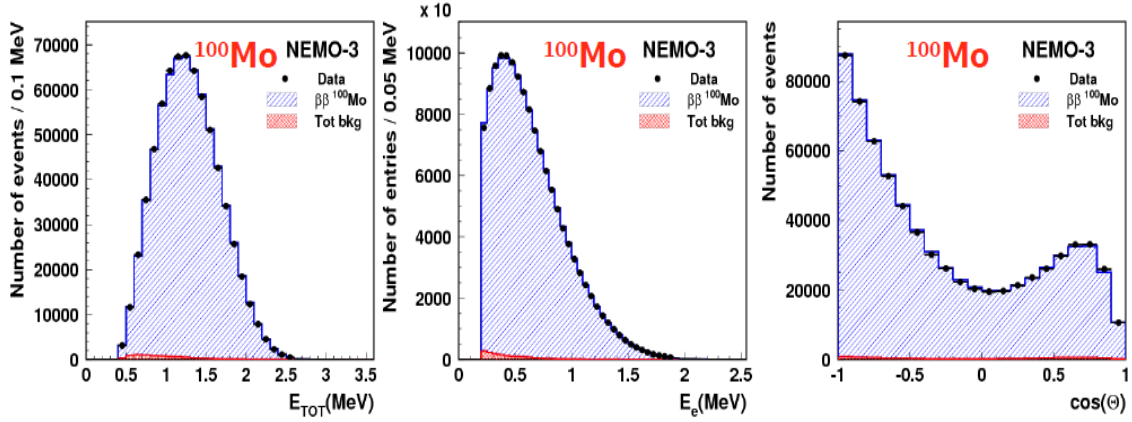}
\caption{Total energy, individual energy and angular distributions
 of the $^{100}$Mo 2$\nu$ events in the NEMO-3 experiment for the 
low radon Phase-II data (4 years) \cite{SIM12}.}
\end{figure}
 
Measurements with the NEMO-3 detector revealed that tracking 
information, combined with time and energy measurements, makes 
it possible to suppress the background efficiently. Using the 
NEMO--3 installation 7 isotopes – $^{100}$Mo (6.9 kg), $^{82}$Se (0.93 kg), 
$^{116}$Cd (405 g), $^{150}$Nd (36.6 g), $^{96}$Zr (9.4 g), $^{130}$Te (454 g) 
and $^{48}$Ca (7 g) are investigated. A full description of the 
detector and its characteristics can be found in \cite{ARN05}.

Figure 2 display the spectrum of $2\nu\beta\beta$   events for $^{100}$Mo 
that were 
collected over 4 years (low radon Phase II). The angular 
distribution and single electron spectrum are also shown. The total 
number of events exceeds 700000 which is much greater than the total 
statistics of all of the preceding experiments with $^{100}$Mo (and even 
greater than the total statistics of all previous $2\nu\beta\beta$ decay 
experiments!). It should also be noted that the background is as 
low as 1.3\% of the total number of $2\nu\beta\beta$   events. Measurements of 
the $2\nu\beta\beta$   decay half-lives have been performed for seven isotopes 
available in NEMO--3. The NEMO--3 results of $2\nu\beta\beta$ half-life measurements 
are given in Table I (only part of full statistic has been analysed). 
For all the isotopes the energy sum spectrum, 
single electron energy spectrum and angular distribution were measured. 
The $^{100}$Mo double beta decay to the 0$^+_1$ excited state of $^{100}$Ru has 
also been measured by NEMO--3 \cite{ARN07}. For $^{100}$Mo, $^{82}$Se, $^{96}$Zr, $^{150}$Nd and 
$^{130}$Te these intermediate results are published. For the other isotopes their 
status is preliminary.

\begin{table*}[h]
\caption{Present results from NEMO--3 (only part of full statistic has been analysed). 
All limits  are presented at a 90\% C.L.}
\vspace{0.5cm}
\begin{tabular}{l|l|l|c}
\hline \multicolumn{1}{c|}{Isotope} &
\multicolumn{1}{c|}{$T_{1/2}(2\nu)$, yr}& \multicolumn{1}{c|}{$T_{1/2}(0\nu)$, yr}&
$T_{1/2}(0\nu\chi^0)$, yr \\
\hline
$^{100}$Mo & $(7.11 \pm 0.02 \pm 0.54)\cdot 10^{18}$ \cite{ARN05a} & $>1.1\times 10^{24}$ \cite{BAR11} 
& $>2.7\times 10^{22}$ \cite{ARN06} \\
$^{100}$Mo--& $(5.7^{+1.3}_{-0.9} \pm 0.8)\cdot10^{20}$ & $>8.9\times 10^{22}$ & - \\
$^{100}$Ru (0$^+_1$)\cite{ARN07} & & & \\
$^{82}$Se& $(9.6 \pm 0.3 \pm 1.0)\cdot10^{19}$ \cite{ARN05a} 
& $>3.6\times 10^{23}$ \cite{BAR11} & $>1.5\times 10^{22}$ \cite{ARN06}  \\
$^{130}$Te \cite{ARN11} & $(7.0 \pm 0.9 \pm 1.1)\cdot10^{20}$ & $>1.3\times 10^{23}$ & $>1.6\times 10^{22}$ \\
$^{150}$Nd \cite{ARG09} & $(9.11^{+0.25}_{-0.22} \pm 0.63)\cdot10^{18}$ 
& $>1.8\times 10^{22}$ & $>1.52\times 10^{21}$ \\
$^{96}$Zr \cite{ARG10} & $(2.35 \pm 0.14 \pm 0.16)\cdot10^{19}$ & $>9.2 \times 10^{21}$ & $>1.9 \times 10^{21}$ \\
$^{116}$Cd & $(2.88 \pm 0.04 \pm 0.16)\cdot10^{19}$ & $>1.3\times 10^{23}$ & - \\
$^{48}$Ca & $(4.4^{+0.5}_{-0.4} \pm 0.4)\cdot10^{19}$ & $>1.3\times 10^{22}$ & - \\
\hline
\end{tabular}

\end{table*}

No evidence for $0\nu\beta\beta$ decay was found for all seven isotopes. 
The associated limits are presented in Table I. Best limit has 
been obtained for $^{100}$Mo (T$_{1/2}^{0\nu} > 1.1\cdot 10^{24} yr$). 
Corresponding limit 
on the neutrino mass is $\langle m_{\nu}\rangle$  $< 0.29-0.7$ eV 
(using nuclear mutrix element (NME) values from 
\cite{KOR07,BAR09,SIM09,RAT10,ROD10}). No evidence 
for decay with Majoron emission ($0\nu\chi^{0}\beta\beta$) was found for all seven 
isotopes too. The limits for $^{100}$Mo, $^{82}$Se, $^{150}$Nd, $^{96}$Zr and $^{130}$Te 
are presented in Table I. In particular, strong limits on "ordinary" 
Majoron (spectral index 1) decay of $^{100}$Mo 
($T_{1/2} > 2.7\cdot10^{22}$ y) and 
$^{82}$Se ($T_{1/2} > 1.5\cdot10^{22}$ y) have been obtained. 
Corresponding bounds 
on the Majoron-neutrino coupling constant have been established, 
$<g_{ee}> < (0.25-0.67) \cdot 10^{-4}$ 
and  $< (0.6-1.9) \cdot 10^{-4}$, respectively (using 
nuclear matrix elements from \cite{KOR07,BAR09,SIM09,RAT10,ROD10,KOR07a,CAU08}).

Data analysis proceeds and Collaboration hope for receiving final 
results for all 7 isotopes in the nearest future (2012--2013).

\subsection{EXO--200 \cite{ACK11,AUG12}}

EXO--200 (Enriched Xenon Observatory) is operating at the Waste Isolation
Pilot Plant (WIPP, 1585 m w.e.) since May 2011. The experiment consists 
of 175 kg of Xe enriched to 80.6\% in $^{136}$Xe housed in a liquid time 
projection chamber (TPC). The TPC is surrounded by passive and active 
shields. This detector measures energy through both ionization and 
scintillation and is capable of effectively rejecting rays through 
topological cuts. EXO--200 has recently claimed the first observation 
of $2\nu\beta\beta$  in $^{136}$Xe (Q$_{\beta\beta}$ = 2458.7 keV) \cite{ACK11}. 
Initial results on $0\nu\beta\beta$ decay together 
with new result for 2$\nu$ mode are published in \cite{AUG12}. The fiducial volume 
used in this analysis contains 79.4 kg of $^{136}$Xe (3.52$\cdot10^{26}$ atoms), 
corresponding to 98.5 kg of active $^{enr}$LXe. Energy resolution is 
10.5 $\%$ (FWHM) at 2.615 MeV using ionization signal only and 4$\%$ (FWHM)  
using both ionization and scintillation signals. Background index (BI) 
in the 0$\nu$ region is $1.4\cdot10^{-3}$ counts/keV$\cdot$kg$\cdot$yr (see Fig. 3). 
Results obtained after 2896.6 h of measurements are the following:
 
\vspace{0.5cm}
$T_{1/2}$ (2$\nu$, $^{136}$Xe) = $[2.23 \pm 0.017 (stat) \pm 0.22 (syst)]\cdot10^{21} yr$ (1)  

$T_{1/2}$ (0$\nu$, $^{136}$Xe) $> 1.6\cdot10^{25} yr$ (90\% C.L.) (2)                                     
\vspace{0.5cm}

Last result provides upper limit $\langle m_{\nu}\rangle$ $<$ 0.14-0.38 eV 
using NME values 
from \cite{KOR07,BAR09,SIM09,ROD10,CAU08}). Taking into account present background one can 
predict that EXO--200 sensitivity after 5 years of data taking will 
be $T_{1/2} \sim 4\cdot10^{25}$ yr ($\langle m_{\nu}\rangle$ $\sim$ 0.09--0.24 eV). 

\begin{figure*}
\begin{center}
\resizebox{0.5\textwidth}{!}{\includegraphics{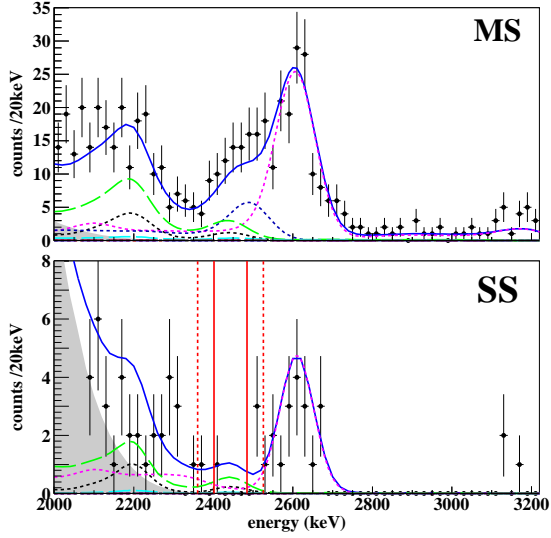}}
\caption{Energy spectra in the $^{136}$Xe Q$_{\beta\beta}$ region for multiple-site (top) 
and single-site (bottom) 
events. The 1 (2) $\sigma$ regions around Q$_{\beta\beta}$ are shown by solid (dashed) vertical lines 
\cite{AUG12}.}
\label{fig3}
\end{center}
\end{figure*}

The project is also a prototype for a planned 1 ton sized experiment 
that may include the
ability to identify the daughter of $^{136}$Ba in real time, effectively 
eliminating all classes of background except that due to 2$\nu$ decay 
(see Section III. E).

\subsection{KamLAND--Zen \cite{GAN12,GAN12a}}

The detector KamLAND--Zen (Fig. 4) is a modification of the existing KamLAND 
detector carried out in the summer of 2011. The $\beta\beta$ source/detector is 13 tons 
of Xe-loaded liquid scintillator (Xe--LS) contained in a 3.08 m diameter 
spherical inner balloon (IB). The IB is constructed from 25 $\mu$m thick 
transparent nylon film and is suspended at the center of the KamLAND 
detector by 12
film straps of the same material. The IB is surrounded by 1 kton of liquid 
scintillator (LS) contained in a 13 m diameter spherical outer balloon (OB) 
made of 135 $\mu$m thick composite film. The outer LS acts as an active shield 
for external $\gamma$'s and as a detector for internal radiation from the Xe--LS or IB. 
The Xe--LS contains (2.52 $\pm$ 0.07) $\%$ by weight of enriched xenon gas (full 
weight of xenon is $\sim$ 
330 kg). The isotopic abundances in the enriched xenon were measured by 
residual gas analyzer to
be (90.93 $\pm$ 0.05) $\%$ of $^{136}$Xe and (8.89 $\pm$ 0.01) $\%$ of $^{134}$Xe. Scintillation light 
is recorded by 1,325 17-inch and 554 20-inch photomultiplier tubes (PMTs). 
Details of the KamLAND detector are given in Ref. \cite{ABE10}. The energy resolution 
at 2.614 MeV is $\sigma$ = (6.6 $\pm$ 0.3)$\%$/$\surd$E (MeV). The vertex resolution is 
$\sigma$ = 15 cm/$\surd$E (MeV). The energy spectrum of $\beta\beta$ decay candidates is shown 
in Fig. 5. Unexpectedly detected background (BI $\approx$ 10$^{-4}$ counts/keV$\cdot$kg$\cdot$yr) 
is $\sim$ two order of magnitude higher than estimated background using previous 
data of KamLAND detector. Nevertheless quite good results for 2$\nu$ decay of 
$^{136}$Xe were obtained. The measured $2\nu\beta\beta$  decay half-life of $^{136}$Xe \cite{GAN12a} is:

\vspace{0.5cm}     
$T_{1/2}$ (2$\nu$, $^{136}$Xe) = $[2.30 \pm 0.02 (stat) \pm 0.12 (syst)]\cdot10^{21}$ yr                                                  (3)
\vspace{0.5cm}

This is consistent with the result obtained by EXO--200 \cite{ACK11,AUG12}. For $0\nu\beta\beta$ decay, 
the data give a
lower limit of $T_{1/2}$ (0$\nu$, $^{136}$Xe) $> 6.2\cdot10^{24}$ yr (90$\%$ C.L.) \cite{GAN12a}, 
which corresponds 
to limit, $\langle m_{\nu}\rangle$ $< 0.22-0.6$ eV using NME values 
from \cite{KOR07,BAR09,SIM09,ROD10,CAU08}. Strong 
limits on neutrinoless double beta decay with emission of Majoron  were 
obtained too (for different modes). In particular, a lower limit on the ordinary Majoron 
 emitting decay (spectral index n = 1) half-life of $^{136}$Xe was obtained 
as $T_{1/2} (0\nu\chi^0$, $^{136}$Xe) $> 2.6\cdot10^{24}$ yr at 90$\%$ C.L. The corresponding upper 
limit on the effective Majoron-neutrino coupling constant, using a range of available 
nuclear matrix calculations, is  $<g_{ee}>$ $<(0.8 - 1.6)\cdot10^{-5}$. This is most 
strong limit on $<g_{ee}>$ from $\beta\beta$ decay experiments.

\begin{figure*}
\begin{center}
\resizebox{0.5\textwidth}{!}{\includegraphics{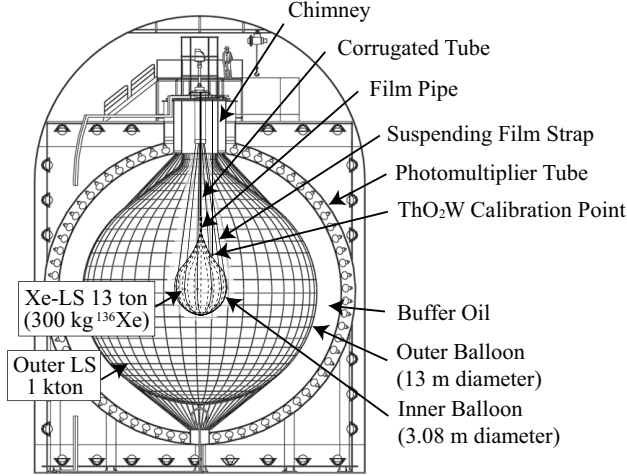}}
\caption{Schematic diagram of the KamLAND--Zen detector \cite{GAN12}.}
\label{fig4}
\end{center}
\end{figure*}

Now the Collaboration undertakes efforts to decrease the background. 
In principle, the background can be lowered approximately in 100 times. 
If it will be done, sensitivity of experiment will essentially increase 
and for 3 years of measurements will be $T_{1/2} \sim 2\cdot10^{26}$ yr that corresponds 
to sensitivity to neutrino mass, $\langle m_{\nu}\rangle$ $\sim 0.04 - 0.11$ eV. After end of the 
first phase of the experiment the phase 2 is planned (see the section III. F).

\begin{figure*}
\begin{center}
\resizebox{0.5\textwidth}{!}{\includegraphics{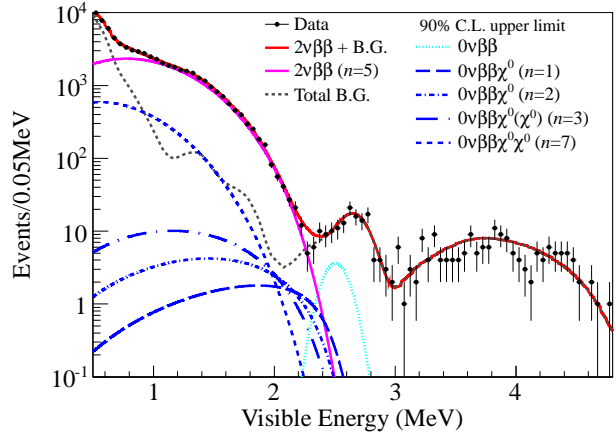}}
\caption{. Energy spectrum of selected $\beta\beta$ decay candidates (data points) together 
with the best-fit backgrounds (gray dashed lines) and $2\nu\beta\beta$  decay (purple solid line), 
and the 90\% C.L. upper limit for $0\nu\beta\beta$ decay and Majoron-emitting 
$0\nu\beta\beta$  decays for each 
spectral index. The red line depicts the sum of the $2\nu\beta\beta$  decay and background spectra. 
Figure is taken from \cite{GAN12a}. }
\label{fig5}
\end{center}
\end{figure*}

\subsection{GERDA-I \cite{CAT12}}

GERDA is a low-background experiment which searches for the neutrinoless double beta
decay of $^{76}$Ge, using an array of high-purity germanium (HPGe) detectors isotopically
enriched in $^{76}$Ge \cite{ABT04}. The detectors are operated naked in ultra radio-pure liquid
argon, which acts as the cooling medium and as a passive shielding against the external
radiation. This innovative design, complemented by a strict material selection for
radio-purity, allows to achieve low background level in the region of the Q-value of
$^{76}$Ge at 2039 keV. The experiment is located in the underground Laboratori Nazionali
del Gran Sasso of the INFN (Italy). The Phase I of GERDA recently has been started using eight
enriched coaxial detectors (totaling approximately 18 kg of $^{76}$Ge). The Phase I comes
after a one-year-long commissioning, in which natural and enriched HPGe detectors
were successfully operated in the GERDA set-up. GERDA-I measurements have been started
in November 2011. Results of first measurement  are presented 
in Fig. 6 (6.1 kg$\cdot$yr of data). 

\begin{figure}
\includegraphics[scale=0.35]{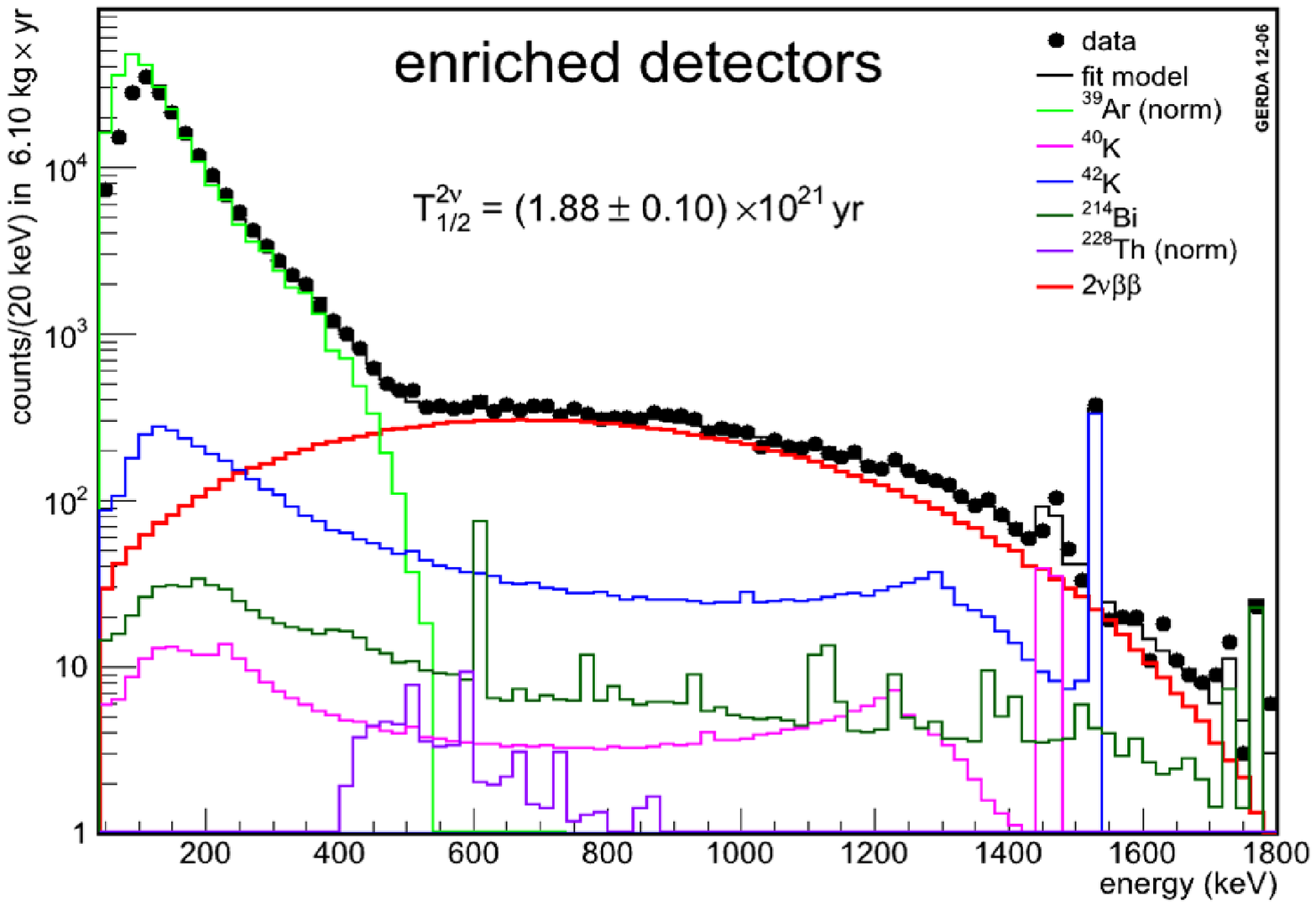}
\includegraphics[scale=0.55]{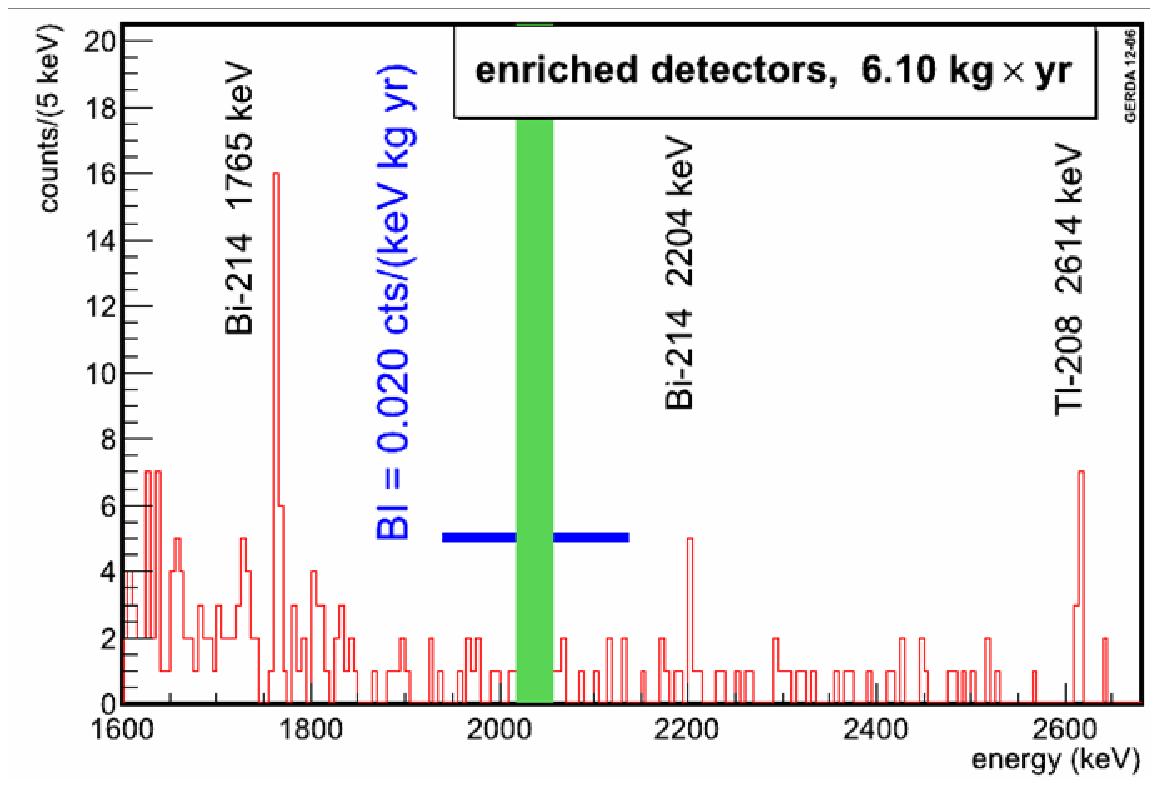}
\caption{First results from GERDA-I (figures are taken from \cite{GRA12}).}
\end{figure}

As can be seen the 2$\nu$ decay contribution is already clearly visible ($T_{1/2}$ (2$\nu$, $^{76}$Ge)
  $\approx 1.88\cdot10^{21}$ yr, preliminary result). Background index in 0$\nu$ region is 
$\sim$ 0.02 c/keV$\cdot$kg$\cdot$yr.
Blind analysis will be applied to the 0$\nu$ region (which is closed now). First result 
will be reported in the end of 2012. Sensitivity of GERDA-I with present background 
is $\sim 2\cdot10^{25}$ yr for one year of measurement. In 2013 new $\sim$ 20 kg of HPGe crystals 
will be added and experiment will be transformed to Phase II (GERDA-II). 
Description of full scale GERDA experiment is done in Sec. III B.

\section{Future large-scale experiments}

Here seven of the most developed and promising experiments which can be realized
 within the next few years are discussed (see Table II). The estimation of the 
sensitivity in the experiments is made using NME from \cite{KOR07,BAR09,SIM09,RAT10,ROD10,KOR07a,CAU08}.

\begin{table}
\caption{Seven most developed and promising projects. 
Sensitivity at 90\% C.L. for one (GERDA-I), three (1-st step of GERDA and MAJORANA, SNO+, and KamLAND--Xe) 
five (EXO, SuperNEMO and CUORE) and ten (full-scale GERDA and MAJORANA) 
years of measurements is presented. M -- mass of isotopes.}
\vspace{0.5cm}
\begin{center}
\begin{tabular}{c|c|c|c|c|c}
\hline
Experiment & Isotope & M, kg & Sensitivity & Sensitivity & Status \\
& &  & $T_{1/2}$, y & $\langle m_{\nu} \rangle$, meV &  \\
\hline
CUORE \cite{GOR12,ARN04} & $^{130}$Te & 200 & $10^{26}$ & 50--130 & in progress \\ 
GERDA \cite{CAT12,ABT04} & $^{76}$Ge & 18 & $2\times10^{25}$ & 200--700 & current \\
&  & 40 & $2\times10^{26}$ & 60--200 & in progress \\
& & 1000 & $6\times10^{27}$ & 10--40 & R\&D\\ 
MAJORANA & $^{76}$Ge & 20--30 & $10^{26}$ & 90--300 & in progress \\
\cite{MAJ03,WIL12}& & 1000 & $6\times10^{27}$ & 10--40 & R\&D \\ 
EXO \cite{ACK11,DAN00} & $^{136}$Xe & $\sim$175 & $4\times10^{25}$ & 90--240 & current \\
& & 1000 & $8\times10^{26}$ & 20--55 & R\&D \\ 
SuperNEMO & $^{82}$Se & 100--200 & (1--2)$\times10^{26}$ & 40--110 & construction of \\
\cite{BAR02,BAR12} & & & & & first module;\\
& & & & & R\&D\\
KamLAND--Zen & $^{136}$Xe & $\sim$330 & 2$\times10^{26}$ & 40--110 & current \\
\cite{GAN12,GAN12a} & & 1000 & $10^{27}$ & 18--50 & R\&D\\
SNO+ \cite{HAR12} & $^{150}$Nd & 50 & 6$\times10^{24}$ & 120--410 & in progress \\
 & & 500 &  3$\times10^{25}$ & 55-180 &  R\&D \\
\hline
\end{tabular}
\end{center}
\end{table}

\subsection{CUORE \cite{GOR12,ARN04}}

This experiment will be run at the Gran Sasso Underground Laboratory (Italy, 3500 m w.e.). 
The plan is to investigate 760 kg of $^{nat}$TeO$_2$, with a total of $\sim$ 200 kg of $^{130}$Te. 
One thousand low temperature ($\sim$8 mK) detectors, each having a weight of 750 g, will 
be manufactured and arranged in 19 towers. One tower is approximately equivalent 
to the CUORICINO detector \cite{AND11}. Planed energy resolution is 5 kev (FWHM). One of 
the problems here is to reduce the background level by a factor of about 15 in 
relation to the background level achieved in the detector CUORICINO. Upon reaching 
a background level of 0.01 c/keV$\cdot$kg$\cdot$yr, the sensitivity of the experiment to 
the 0$\nu$ decay of $^{130}$Te for 5 y of measurements and at 90$\%$ C.L. will become approximately 
10$^{26}$ yr ($\langle m_{\nu}\rangle$ $\sim$ 0.05-0.13 eV) - see discussion in \cite{ALE11}. 
The experiment has been 
approved and funded. A general test of the CUORE detector, comprising a single 
tower and named CUORE--0, will start to take data in 2012. The full-scale 
CUORE will start in $\sim$ 2014.

\subsection{GERDA \cite{CAT12,ABT04}}

This is one of two planned experiments with $^{76}$Ge (along with the MAJORANA experiment). 
The experiment is to be located in the Gran Sasso Underground Laboratory (Italy, 3500 m w.e.). 
The proposal is based on ideas and approaches which were proposed for GENIUS \cite{KLA98} and the 
GEM \cite{ZDE01} experiments. The idea is to place "naked" HPGe detectors in highly purified liquid
argon (as passive and active shield). It minimizes the weight of construction material 
near the detectors and decreases the level of background. The liquid argon dewar is 
placed into a vessel of very pure water. The water plays a role of passive and active 
(Cherenkov radiation) shield. The proposal involves three phases. In the first phase, 
the existing HPGe detectors ($\sim$ 18 kg), which previously were used in the Heidelberg-Moscow 
\cite{KLA01} and IGEX \cite{AAL02} experiments, are utilized (see Sec. II. D). In the second phase $\sim$ 40 kg 
of enriched Ge will be investigated. In the third phase the plan is to use $\sim$ 1000 kg 
of $^{76}$Ge. The sensitivity of the second phase (for three years of measurement) will be 
$T_{1/2} \sim 2\cdot10^{26}$ yr. This corresponds to a sensitivity for $\langle m_{\nu}\rangle$ 
at the level of $\sim$ 0.06--0.2 eV.

The first two phases have been approved and funded. First phase will be finished in the 
end of 2012. The second phase setup is in an advanced construction stage and data taking 
is foreseen for 2013. The results of these steps will play an important role in the 
decision to support the full scale experiment.

\subsection{MAJORANA \cite{MAJ03,WIL12}}

The MAJORANA facility will consist of $\sim$ 1000 HPGe detectors manufactured from enriched 
germanium (the enrichment is $> 86\%$). The total mass of enriched germanium will be 1000 kg. 
The facility is designed in such a way that it will consist of many individual 
supercryostats manufactured from low radioactive copper, each containing HPGe detectors. 
The entire facility will be surrounded by a passive shield and will be located at 
an underground laboratory in the United States. Only the total energy deposition will
 be utilized in measuring the $0\nu\beta\beta$ decay of $^{76}$Ge to the ground state of the daughter 
nucleus. The use of HPGe detectors, pulse shape analysis, anticoincidence, and low 
radioactivity structural materials will make it possible to reduce the background to 
a value below 2.5$\cdot10^{-4}$ c/keV$\cdot$kg$\cdot$yr and to reach a sensitivity of 
about $6\cdot10^{27}$ y 
within ten years of measurements. The corresponding sensitivity to the effective 
mass of the Majorana neutrino is about 0.01 to 0.04 eV. The measurement of the 
$0\nu\beta\beta$ decay of $^{76}$Ge to the 0$^+$ excited state of the daughter nucleus will be performed 
by recording two cascade photons and two beta electrons. The planned sensitivity for 
this process is about 10$^{27}$ y. In the first step $\sim$ 20--30 kg of $^{76}$Ge will be investigated 
(MAJORANA Demonstrator). 
It is anticipated that the sensitivity to $0\nu\beta\beta$ decay to the ground state of the daughter 
nuclei for 3 years of measurement will be $T_{1/2} \sim 10^{26}$ yr. It will reject or confirm 
the "positive" result from \cite{KLA01a,KLA04,KLA06}. 
Sensitivity to $\langle m_{\nu}\rangle$ will be $\sim$ 0.09--0.3 eV. 
During 
this time different methods and technical questions will be checked and possible 
background problems will be investigated. The MAJORANA Demonstrator is being constructed at
the Sanford Underground Research Facility (SURF) at the old Homestake gold mine
in Lead, SD. The first cryostat is planned for commissioning in 2013.

The Majorana and GERDA collaborations are cooperating in efforts to design
a large-mass ($\sim$ 1 ton) Ge detector. The confguration of such an experiment will be optimized
based on the outcomes of the MAJORANA Demonstrator and GERDA Phase-II.

\subsection{SuperNEMO \cite{BAR02,BAR12}}

The NEMO Collaboration has studied and is pursing an experiment that will observe 
100--200 kg of $^{82}$Se with the aim of reaching a sensitivity for the 0$\nu$ decay mode 
at the level of $T_{1/2} \sim 
(1-2)\cdot10^{26}$ y. The corresponding sensitivity to the neutrino mass is 0.04 to 0.11 eV. 
In order to accomplish this goal, it is proposed to use the experimental procedures 
nearly identical to that in the NEMO--3 experiment (see Sec. II. A). The new detector 
will have planar geometry and will consist of 20 identical modules (7 kg of $^{82}$Se 
in each sector). A $^{82}$Se source having a thickness of about 40 mg/cm$^2$ and a very low 
content of radioactive admixtures is placed at the center of the modules. The detector 
will again record all features of double beta decay: the electron energy will be 
recorded by counters based on plastic scintillators ($\Delta$E/E $\sim 8\%$ (FWHM) at E = 1 MeV), 
while tracks will be reconstructed with the aid of Geiger counters. The same device 
can be used to investigate $^{150}$Nd, $^{100}$Mo, $^{116}$Cd, and $^{130}$Te 
with a sensitivity to $0\nu\beta\beta$ 
decay at a level of about $(0.5-1)\cdot10^{26}$ yr \cite{BAR02}. The use of an already tested 
experimental technique is an appealing feature of this experiment. The plan is to arrange 
the equipment at the new Frejus Underground Laboratory (France; 4800 m w.e.). The 
construction and commissioning of the Demonstrator (first module) will be completed in 2013--2014.

\subsection{EXO \cite{ACK11,DAN00}}

In this experiment the plan is to implement Moe's proposal of 1991 \cite{MOE91}. Specifically 
it is to record both ionization electrons and the Ba$^+$ ion originating from the 
double beta decay process $^{136}$Xe $\to$ $^{136}$Ba$^{++}$ + 2e$^-$. 
In \cite{DAN00}, it is proposed to operate 
with 1t of $^{136}$Xe. The actual technical implementation of the experiment has not yet 
been developed. One of the possible schemes is to fill a TPC with liquid enriched 
xenon. To avoid the background from the 2$\nu$ decay of $^{136}$Xe, the energy resolution 
of the detector must not be poorer than 3.8\% (FWHM) at an energy of 2.5 MeV 
(ionization and scintillation signals will be detected). In the 0$\nu$ decay of $^{136}$Xe, 
the TPC will measure the energy of two electrons and the coordinates of the event
to within a few millimeters. After that, using a special stick Ba ions will be 
removed from the liquid and then will be registered in a special cell by resonance 
excitation. For Ba$^{++}$ to undergo a transition to a state of Ba$^+$, a special gas is 
added to xenon. The authors of the project assume that the background will be 
reduced to one event within five years of measurements. Given a 70\% detection 
efficiency it will be possible to reach a sensitivity of about 8$\cdot10^{26}$ yr for 
the $^{136}$Xe half-life and a sensitivity of about 0.02 to 0.06 eV to the neutrino 
mass. One should note that the principle difficulty in this experiment is 
associated with detecting the Ba$^+$ ion with a reasonably high efficiency. 
This issue calls for thorough experimental tests, and positive results have 
yet to be obtained. As the first stage of the experiment EXO--200 use 175 kg 
of $^{136}$Xe without Ba ion identification (see Sec. II. B). 

\subsection{KamLAND-Zen--2}

KamLAND--Zen is an upgrade of the KamLAND setup \cite{ABE10}. The idea is to convert 
it to neutrinoless double beta decay search by dissolving Xe gas in the liquid scintillator. 
This approach was proposed by R. Raghavan in 1994 \cite{RAG94}. At the first step this mixture 
(330 kg of Xe in 13 tons of liquid scintillator) will be contained in a small balloon 
suspended in the centre of the KamLAND sphere. It will guarantee low background level 
and high sensitivity of the experiment. This experiment (KamLAND--Zen) is in a stage of a 
data taking and some more years will proceed (see Sec. II. C). Experiment KamLAND-Zen--2 
with 1000 kg of the enriched xenon will be the next step. It is planned to upgrade the 
existing detector. It is supposed that in the new inner balloon more bright liquid 
scintillator will be used and the number of PMTs will be increased. All this will 
allow to improve essentially energy resolution of the detector and, thereby, to 
increase sensitivity of experiment to double beta decay (see Table II). 
KamLAND-Zen--2 will start after $\sim$ 2015. 

\subsection{SNO+ \cite{HAR12}}

SNO+ is an upgrade of the solar neutrino experiment SNO (Canada), aiming at 
filling the SNO detector with Nd loaded liquid scintillator to investigate the 
isotope $^{150}$Nd. With 0.1\% loading SNO+ will use 0.78 tons of neodymium and contain 
43.7 kg of $^{150}$Nd with no enrichment. SNO+ is in construction phase with natural neodymium. 
Data taking is foreseen in 2013--2014. After 3 yr of data taking sensitivity will 
be $\sim 6\cdot10^{24}$ yr (or 0.12--0.41 eV for $\langle m_{\nu}\rangle$). 
Finally 500 kg of enriched $^{150}$Nd will 
be used (if enrichment of such quantity of Nd will be possible). Planned sensitivity
 is $\sim 3\cdot10^{25}$ yr (or 0.055--0.18 eV for $\langle m_{\nu}\rangle$).

\begin{table*}[h]
\caption{Best current results concerning the search for $0\nu\beta\beta$ decay. 
All bounds are given with 90\% C.L. The bounds on the effective mass of 
the Majorana neutrino $\langle m_{\nu}\rangle$ were obtained using the calculated nuclear 
matrix elements from \cite{KOR07,BAR09,SIM09,RAT10,ROD10,KOR07a,CAU08}.}
\vspace{0.5cm}
\begin{tabular}{l|l|l|c}
\hline \multicolumn{1}{c|}{Isotope} &
\multicolumn{1}{c|}{$E_{2\beta }$, keV}& \multicolumn{1}{c|}{$T_{1
/ 2}$ , yr}&
$\langle m_{\nu }\rangle $, eV \\
\hline
$^{48}$Ca & 4272 & $>5.8\times 10^{22}$ \cite{UME08}&$<14$ \\
{ $^{76}$Ge} & 2039.0& $>1.9\times 10^{25}$ \cite{KLA01}& ${<0.20-0.69}$ \\
$^{82}$Se & 2996 & $>3.6\times 10^{23}$ \cite{BAR11}&$<0.77-2.4$\phantom{0} \\
$^{96}$Zr & 3350 & $>9.2\times 10^{21 }$ \cite{ARG10} &\phantom{0}$<3.9-13.7$ \\
{ $^{100}$Mo} & 3034.4 & $>1.1\times 10^{24 }$ \cite{BAR11}& ${ <0.29-0.70}$ \\
$^{116}$Cd & 2813.5 & $>1.7 \times 10^{23}$ \cite{DAN03}&$<1.16-2.16$ \\
$^{128}$Te & 867 & $>1.5\times 10^{24}$~(geochemistry) (\cite{MAN01,BAR10}) &$<1.8-4.2$ \\
{ $^{130}$Te} & 2527.5 &\phantom{0.}$>2.8\times 10^{24}$ \cite{AND11}& ${ <0.35-0.77}$ \\
{ $^{136}$Xe} & 2458.7 & $>1.6\times 10^{25}$ \cite{AUG12}& ${ <0.14-0.38}$ \\
$^{150}$Nd & 3371.4 & $>1.8\times 10^{22}$ \cite{ARG09}&$<2.2-7.5$ \\
\hline
\end{tabular}

\end{table*}

\section{Conclusion}

Best present limits on $0\nu\beta\beta$  decay and on $\langle m_{\nu}\rangle$ are presented in Table III.  
It is visible that the most strong limits are received in experiments with $^{136}$Xe, 
$^{76}$Ge, $^{100}$Mo and $^{130}$Te. Considering existing uncertainty in values of NME 
it is possible to obtain conservative limit { $\langle m_{\nu}\rangle$ $<$ 0.4 eV} 
(using conservative EXO-200 value). It is possible to
expect that in the next few years sensitivity to $\langle m_{\nu}\rangle$ will be improved by 
efforts of experiments of EXO--200, KamLAND--Zen, GERDA--II, MAJORANA--Demonstrator, 
CUORE--0 several times and will reach values $\sim$ 0.1--0.3 eV. Start of full-scale 
experiments will allow to reach in 2015--2020 sensitivity to $\langle m_{\nu}\rangle$ at the level 
0.01--0.1 eV that will allow to begin testing of inverted hierarchy region ($\sim$ 50 meV). 
Using modern experimental approaches it will be extremely difficult to reach 
sensitivity to $\langle m_{\nu}\rangle$ on the level of $\sim$ 3--5 meV (normal hierarchy region). For 
this purpose it is required to increase mass of a studied isotope to $\sim 10$ tons 
and to provide almost zero level of a background in studied area. Nevertheless 
it was shown, what even using known today methods such possibility, in principle, 
exists (see \cite{BAR12a}).


\begin{thebibliography}{References}

\bibitem{ELL87} 
Elliott S. R., Hahn A. A., Moe M. K., Phys. Rev. Lett. 59 (1987) 2020.
\bibitem{KLA01}
Klapdor-Kleingrothaus H. V. et al., Eur. Phys. J. A 12 (2001) 147.
\bibitem{AAL02} 
Aalseth  C. E. et al., Phys. Rev. D 65 (2002) 092007.
\bibitem{ARN05} 
Arnold R. et al., Nucl. Instr. Meth. A 536 (2005) 79.
\bibitem{AND11} 
Andreotti E., Astropart. Phys. 34 (2011) 822.
\bibitem{ACK11} 
Ackerman N. et al., Phys. Rev. Lett. 107 (2011) 212501.
\bibitem{GAN12} 
Gando A. et al., Phys. Rev. C 85 (2012) 045504.
\bibitem{HAR12} 
Hartnell J., J. Phys. Conf. Ser. 375 (2012) 042015.
\bibitem{GOR12} 
Gorla P., J. Phys. Conf. Ser. 375 (2012) 042013.
\bibitem{SIM12}
Simard L., J. Phys. Conf. Ser. 375 (2012) 042011. 
\bibitem{ARN05a} 
Arnold R. et al., Phys. Rev. Lett. 95 (2005) 182302.
\bibitem{BAR11} 
Barabash A. S., and Brudanin V. B., Phys. At. Nucl. 74 (2011) 312.
\bibitem{ARN06} 
Arnold R. et al., Nucl. Phys. A 765 (2006) 483.
\bibitem{ARN07} 
Arnold R. et al., Nucl. Phys. A 781 (2007) 209.
\bibitem{ARN11} 
Arnold R. et al., Phys. Rev. Lett. 107 (2011) 062504.
\bibitem{ARG09} 
Argyriades J. et al., Phys. Rev. C 80 (2009) 032501R.
\bibitem{ARG10} 
Argyriades J. et al., Nucl. Phys. A 847 (2010) 168.
\bibitem{KOR07} 
Kortelainen M., and Suhonen J., Phys. Rev. C 76 (2007) 024315.
\bibitem{BAR09} 
Barea J. and Iachello F., Phys. Rev. C 79 (2009) 044301.
\bibitem{SIM09} 
Simkovic F. et al., Phys. Rev. C 79 (2009) 055501.
\bibitem{RAT10} 
Rath P. K. et al., Phys. Rev. C 82 (2010) 064310.
\bibitem{ROD10} 
Rodrigues T. R., and Martinez-Pinedo G. M., Phys. Rev. Lett. 105 (2010) 252503.
\bibitem{KOR07a} 
Kortelainen M., and Suhonen J., Phys. Rev. C 75 (2007) 051303R.
\bibitem{CAU08} 
Caurier E. et al., Phys. Rev. Lett. 100 (2008) 052503.
\bibitem{AUG12} 
Auger M. et al., arXiv:hep-ex/1205.5608.
\bibitem{GAN12a} 
Gando A. et al., Phys. Rev. C 86 (2012) 021601R.
\bibitem{ABE10} 
Abe S. et al., Phys. Rev. C 81 (2010) 025807.
\bibitem{CAT12} 
Cattadori C. M., J. Phys. Conf. Ser. 375 (2012) 042008.
\bibitem{ABT04} 
Abt I. et al., arXiv:hep-ex/0404039.
\bibitem{GRA12}
Grabmayr P., talk at Int. Conf. Neutrino'2012 (Kyoto, Japan), June 4-9, 2012.
\bibitem{ARN04} 
Arnaboldi C. et al., Nucl. Inst. Meth. A 518 (2004) 775.
\bibitem{MAJ03} 
Majorana Collaboration, arXiv:nucl-ex/0311013.
\bibitem{WIL12} 
Wilkerson J. F. et al., J. Phys. Conf. Ser. 375 (2012) 042010.
\bibitem{BAR02} 
Barabash A. S., Czech. J. Phys. 52 (2002) 575.
\bibitem{BAR12} 
Barabash A. S., J. Phys. Conf. Ser. 375 (2012) 042012.
\bibitem{DAN00} 
Danilov M. et al., Phys. Lett. B 480 (2000) 12.
\bibitem{ALE11} 
Alessandria F. et al., arXiv:nucl-ex/1109.0494.
\bibitem{KLA98} 
Klapdor-Kleingrothaus H. V., Hellmig J, and Hirsch M., J. Phys. G 24 (1998) 483.
\bibitem{ZDE01} 
Zdesenko Yu. G., Ponkratenko O. A., and Tretyak V. I., J. Phys. G 27 (2001) 2129.
\bibitem{KLA01a} 
Klapdor-Kleingrothaus H. V. et al., Mod. Phys. Lett. A 16 (2001) 2409.
\bibitem{KLA04} 
Klapdor-Kleingrothaus H. V. et al., Phys. Lett. B 586 (2004) 198.
\bibitem{KLA06} 
Klapdor-Kleingrothaus H. V. et al., Mod. Phys. Lett. A 21 (2006) 1547.
\bibitem{MOE91} 
Moe M., Phys. Rev. C 44 (1991) R931.
\bibitem{RAG94} 
Raghavan R. S., Phys. Rev. Lett. 72 (1994) 1411.
\bibitem{UME08} 
Umehara S. et al., Phys. Rev. C 78 (2008) 058501.
\bibitem{DAN03} 
Danevich F. A. et al., Phys. Rev. C 68 (2003) 035501.
\bibitem{MAN01}
Manuel O.K., in Proc. Int. Symp. "Nuclear Beta Decay  and Neutrino
(Osaka'86)", (World Scientific, Singapore, 1986), p. 71.
\bibitem{BAR10}
Barabash A.S., Phys. Rev. C 81 (2010) 035501.
\bibitem{BAR12a} 
Barabash A. S., J. Phys. G 39 (2012) 085103.




\end{thebibliography}
\end{document}